\documentclass[twocolumn,prl,showpacs,amsfonts,amsmath,assymb,eufrak]{revtex4-1}
\usepackage{epsfig}
\usepackage{color}
\usepackage{bm}
\usepackage{braket}
\usepackage{graphicx}
\usepackage{amsthm}

\usepackage{calc}

\newcommand{\eq}[1]{\begin{equation} #1 \end{equation}}
\newcommand{\eqa}[2]{\begin{equation} #1 \label{#2} \end{equation}}
\newcommand{\balign}[1]{\begin{align} #1 \end{align}}

\newcommand{\bs}{\boldsymbol}

\newcommand{\figin}[4]
{\begin{figure}[tb]
\centering
\includegraphics[width= #1]{#2.pdf}
\caption{#3}
\label{f:#4}
\end{figure}}

\newcommand{\todayd}{\the\year/\the\month/\the\day}

\newcommand{\bib}{\bibitem}

\newcommand{\lr}{\leftrightarrow}

\newcommand{\lmd}{\lambda}

\newcommand{\lb}{\label}
\newcommand{\nt}{\notag}

\newcommand{\bel}{\begin{easylist}}
\newcommand{\eel}{\end{easylist}}

\newcommand{\eref}[1]{Eq.~\eqref{#1}}

\newcommand{\fref}[1]{Fig.~\ref{f:#1}}

\def \({\left(}
\def \){\right)}
\def \[{\left[}
\def \]{\right]}
\newcommand{\la}{\left\langle}
\newcommand{\ra}{\right\rangle}
\newcommand{\abs}[1]{\left|#1\right|}

\newcommand{\sumtwo}[2]%
{\mathop{\sum_{#1}}_{#2}}
\newcommand{\sumthree}[3]%
{\mathop{\mathop{\sum_{#1}}_{#2}}_{#3}}
\newcommand{\sumfour}[4]%
{\mathop{\mathop{\mathop{\sum_{#1}}_{#2}}_{#3}}_{#4}} 
\newcommand{\prodtwo}[2]%
{\mathop{\prod_{#1}}_{#2}}
\newcommand{\mintwo}[2]%
{\mathop{\min_{#1}}_{#2}}
\newcommand{\maxtwo}[2]%
{\mathop{\max_{#1}}_{#2}}
\newcommand{\maxthree}[3]%
{\mathop{\mathop{\max_{#1}}_{#2}}_{#3}}
\newcommand{\limtwo}[2]%
{\mathop{\lim_{#1}}_{#2}}
\newcommand{\suptwo}[2]%
{\mathop{\sup_{#1}}_{#2}}
\newcommand{\supthree}[3]%
{\mathop{\mathop{\sup_{#1}}_{#2}}_{#3}}
\newcommand{\supfour}[4]%
{\mathop{\mathop{\mathop{\sup_{#1}}_{#2}}_{#3}}_{#4}} 
\newcommand{\inftwo}[2]%
{\mathop{\inf_{#1}}_{#2}}
\newcommand{\infthree}[3]%
{\mathop{\mathop{\inf_{#1}}_{#2}}_{#3}}
\newcommand{\inffour}[4]%
{\mathop{\mathop{\mathop{\inf_{#1}}_{#2}}_{#3}}_{#4}} 

\newcommand\calC{{\cal C}}

\newcommand{\bsv}{\boldsymbol{v}}

\newcommand{\bsX}{\boldsymbol{X}}

\newcommand{\Di}{\mathit{\Delta}}

\newcommand{\para}[1]{{\em #1}\/.---}

\def\rnum#1{\resizebox{0.5em}{\height}{\expandafter{\romannumeral #1}}}
\def\Rnum#1{\resizebox{0.5em}{\height}{\uppercase\expandafter{\romannumeral #1}}}

\newcommand{\pss}{p^{\rm ss}}
\newcommand{\Jss}{J^{\rm ss}}
\newcommand{\dsgm}{\dot{\sigma}}
\newcommand{\vos}{v_{\rm osci}}
\newcommand{\wos}{w_{\rm osci}}
\newcommand{\dPi}{\dot{\Pi}}

\makeatletter
\renewcommand{\@cite}[1]{\textsuperscript{#1)}}
\makeatother

\begin{document}

\preprint{APS/123-QED}

\newcommand{\titlehere}{Entropy production limits all fluctuation oscillations}

\newcommand{\titlename}{\titlehere}

\preprint{APS/123-QED}

\title{\titlename}

\author{Naoto Shiraishi}
\affiliation{Department of Physics, Gakushuin University, 1-5-1 Mejiro, Toshima-ku, Tokyo 171-8588, Japan}

\date{\today}

\begin{abstract}
The oscillation of fluctuation with two state observables is investigated.
Following the idea of Ohga, {\it et al.} [arXiv:2303.13116], we find that the fluctuation oscillation relative to their autocorrelations is bounded from above by the entropy production per characteristic maximum oscillation time.
Our result applies to a variety of systems including Langevin systems, chemical reaction systems, and macroscopic systems.
In addition, our bound consists of experimentally tractable quantities, which enables us to examine our inequality experimentally.
\end{abstract}

\maketitle


\para{Introduction}
Entropy production plays a pivotal role in nonequilibrium statistical mechanics, quantifying the degree of thermodynamic irreversibility of processes.
The celebrated fluctuation theorem~\cite{ECM93, GC95, Kur98, Mae99, Jar00} and its variants~\cite{Jar97, HS01, SU12, SS15, PE17} clearly show the mathematical structure of the thermodynamic irreversibility in an equality form.
In addition, entropy production satisfies not only equalities but also various inequalities as upper bounds of quantities~\cite{Shibook}.
One famous example is the classical speed limit inequalities~\cite{SFS18, VVH20, Dec22, VS23} and the trade-off relation between efficiency and power of heat engines~\cite{SST16, SS19}.
Here, entropy production bounds the speed of processes: a quick process should accompany much entropy production.
The thermodynamic uncertainty relation~\cite{BS15, Ging16, GRH17, DS20, Shi21, DS21, DG23} is another example, where entropy production bounds the relative fluctuation of general currents.
Furthermore, entropy production serves as a restriction on possible paths in state space observed in relaxation processes~\cite{SS19b, Kol22}.

We investigate this direction further in the context of oscillation phenomena.
Our main subject in this Letter is the stationary fluctuation oscillation with two state variables $a$ and $b$ defined as
\eq{
\alpha_{ab}:=\frac12 \la a\dot{b}-b\dot{a}\ra ,
}
which is also called as {\it irreversible circulation of fluctuation}~\cite{TT74} and {\it asymmetric of cross-correlation}~\cite{OIK23} (see \fref{area}).
Since the stationary fluctuation oscillation takes a nonzero value only at nonequilibrium stationary states, this quantity is sometimes regarded as the measure of nonequilibriumness.

Oscillation phenomena including chemical oscillations were investigated in the field of nonlinear physics~\cite{GPbook, TOT74, Kurbook}, and have attracted renewed interest from the viewpoint of stochastic thermodynamics~\cite{Cao15, BS17, HTE18, Fei18, JV20, KG20, Zha20, CLT21, Raz21, OSB22, KOI23}.
Recently, interesting progress was provided by Ohga, {\it et al.}~\cite{OIK23}, which proposes a bound on fluctuation oscillation relative to autocorrelation by using the maximum cycle affinity in the transition network.
This bound builds a bridge between fluctuation oscillation and some thermodynamic quantity.
However, the maximum cycle affinity is not easy to measure in experiments, especially for mesoscopic and macroscopic systems.
In particular, the connection to entropy production has not yet been addressed.

In this short Letter, we prove the upper bounds on fluctuation oscillation by entropy production per characteristic time length of oscillation.
Our result clearly shows that a long-lived oscillation inevitably accompanies much dissipation.
Employing the geometric interpretation proposed by Ohga, {\it et al.}~\cite{OIK23}, we can derive our results transparently.
Our result has wide applicability from particle systems in continuous space, chemical reaction systems, to macroscopic systems.
Another advantage of our result is that the inequality consists only of fluctuation oscillation, autocorrelation, the operator norm of observables, and entropy production rate, all of which are tractable in experiments.
Thus, our relation serves as a good stage to test thermodynamic properties in oscillation phenomena.

\figin{8cm}{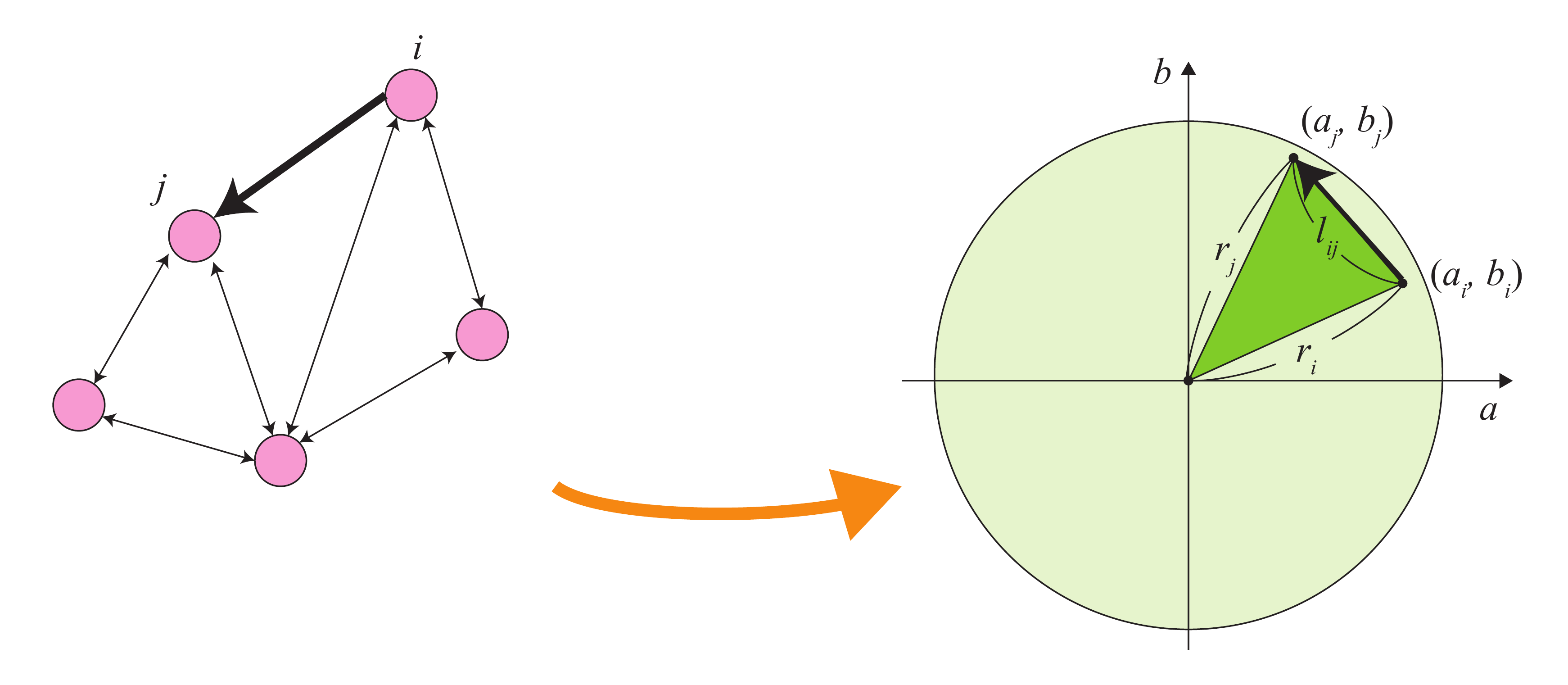}{
An example of state space and its geometric interpretation of fluctuation oscillation and other quantities in the $a$-$b$ space.
The terms in the fluctuation oscillation $\alpha_{ab}=\langle a\dot{b}-b\dot{a}\rangle$ can be interpreted as the area with dark green.
We also express $\ell_{ij}=\sqrt{(a_i-a_j)^2+(b_i-b_j)^2}$ and $r_i=\sqrt{a_i^2+b_i^2}$.
All points $(a_i,b_i)$ are in the circle with diameter $\|\sqrt{a^2+b^2}\|$ drawn in light green.
}{area}

\para{Setup and main result}
We consider a Markov jump process on discrete states, whose time evolution is given by the following master equation:
\eq{
\frac{d}{dt}p_i=\sum_j R_{ij}p_j.
}
Here, $p_i$ is the probability distribution of state $i$, and $R$ is a transition matrix satisfying nonnegativity $R_{ij}\geq 0$ ($i\neq j$) and normalization condition $\sum_i R_{ij}=0$.
We assume the local detailed-balance condition, with which the stationary entropy production rate $\dsgm$ is expressed as
\eq{
\dsgm=\sum_{i,j}R_{ij}\pss_j \ln \frac{R_{ij}\pss_j}{R_{ji}\pss_i}
}
with the stationary distribution $\pss$.

The stationary fluctuation oscillation with $a$ and $b$ can be expressed as
\eq{
\alpha_{ab}=\frac12 \sum_{i,j} (a_jb_i-a_ib_j) R_{ij}\pss_j=\frac12 \sum_{(i,j)} (a_jb_i-a_ib_j)\Jss_{ij},
}
where $\Jss_{ij}:=R_{ij}\pss_j-R_{ji}\pss_i$ is the stationary current between $j$ and $i$, and $\sum_{(i,j)}$ represents the sum over a pair of $i$ and $j$ (i.e., we take only one of $ij$ and $ji$).
We introduce the autocorrelation of $a$ defined as
\eq{
D_a:=-\la a\dot{a}\ra=\frac12\sum_{i,j}(a_i-a_j)^2R_{ij}\pss_j,
}
and normalize the fluctuation oscillation $\alpha_{ab}$ by the average of autocorrelations of $a$ and $b$; $(D_a+D_b)/2$.

Below we present two upper bounds on the normalized fluctuation oscillation $2\alpha_{ab}/(D_a+D_b)$ with the stationary entropy production rate $\dsgm$ relative to the maximum speed of oscillation.
Two inequalities employ different measures of the speed of oscillation.
In the first inequality, we characterize the speed of oscillation by the fluctuation oscillation divided by the area of the circle with a diameter equal to the maximum of $\sqrt{a^2+b^2}$.
The obtained bound is
\eqa{
\frac{2\abs{\alpha_{ab}}}{D_a+D_b}\leq \frac{\dsgm}{2\pi \wos},
}{main-1}
which is our first main result.
Here, $\wos$ is defined as
\eq{
\wos:=\frac{\abs{\alpha_{ab}}}{\pi \|a^2+b^2\|}
}
with the operator norm $\|a^2+b^2\|=\max_i [a^2_i+b^2_i]$.
The denominator represents the area of the circle with diameter $\|\sqrt{a^2+b^2}\|$.
Since $\alpha_{ab}$ can be regarded as the area of a circular sector (see \fref{area}) with a diameter less than $\|\sqrt{a^2+b^2}\|$, we can see $\wos$ as the maximum angular velocity.
We remark that in the definition of $\wos$ we can replace the position of the origin to $(a', b')$ and define as $\wos=\abs{\alpha_{ab}}/\pi \| (a-a')^2+(b-b')^2\|$, with which we can derive the same bound \eqref{main-1}.

The idea behind the second inequality is closer to that shown in Ohga, {\it et al}.~\cite{OIK23}, which heavily relies on the isoperimetric inequality.
From the viewpoint of the isoperimetric inequality, the area of a circle is connected to the square of the perimeter of the circle.
In this line, we claim our second main result:
\eqa{
\frac{2\abs{\alpha_{ab}}}{D_a+D_b}\leq  \frac{\dsgm}{2\pi\vos}.
}{main-2}
Here, $\vos$ is defined as
\eq{
\vos:=\frac{4\pi \abs{\alpha_{ab}}}{(\max_{C\in \calC_{\rm ucd}}\ell_C)^2},
}
where $\ell_C:=\sum_{(i,j)\in C} \sqrt{(a_i-a_j)^2+(b_i-b_j)^2}$ is the length of cycle $C$ in the $a$-$b$ plane, and $\calC_{\rm ucd}$ is a set of cycles in the uniform cycle decomposition~\cite{Sch76}.
The uniform cycle decomposition is a cycle decomposition such that the direction of cycles and that of current coincide on any edge, whose existence is established.
As mentioned above, $\ell_C^2/4\pi$ corresponds to the area bounded by $C$.

These two inequalities clearly show that possible fluctuation oscillation is bounded above by the dissipation.
Some arguments shown in the remainder suggest that the first inequality \eqref{main-1} is more useful in several places than the second one \eqref{main-2}.
We remark that the denominators of $\wos$ and $\vos$ depend only on the state variables $a$ and $b$ and topology of the transition map, and are independent of the transition rates and the stationary distribution (except for the determination of the uniform cycle decomposition in the case of \eqref{main-2}).

\para{Remarks on obtained inequalities}
Before going to the proof of these bounds, we here discuss their physical implications and compare our bounds with other relevant works.

We first remark on the connection to the response theory.
Consider a macroscopic stationary system and apply the system size expansion to it.
In the lowest order, the fluctuation of observables in the vector form $\bsX=(\Di a, \Di b)^\top$ obey a stochastic equation $\dot{\bsX}=L\bsX+{\bs \xi}$, where $L$ is the response matrix and ${\bs \xi}$ is the noise term.
In a nonequilibrium stationary state, the Onsager reciprocity relation no longer holds, $L_{12}\neq L_{21}$, and its discrepancy is known to be equal to the fluctuation oscillation (the irreversible circulation of fluctuation): $(L_{ab}-L_{ba})/2=\alpha_{ab}$~\cite{TT74}.
Thus, our results also serve as a bound on the antisymmetric part of the response matrix by entropy production, which is sometimes referred to as a measure of the degree of nonequilibriumness.

A further implication is seen in the bifurcation phenomena.
When bifurcation occurs, the fluctuation of an observable ($D_a$ or $D_b$) diverges, and in some cases, the oscillation fluctuation ($\alpha_{ab}$) also diverges simultaneously, which are called soft-mode instability and hard-mode instability, respectively~\cite{TOT74}.
Since the entropy production rate $\dsgm$ does not diverge at the bifurcation point, our results can also be read as a bound on the speed of divergence of these two instabilities in terms of entropy production.

Next, we compare our results with the conjecture raised by Oberreiter, {\it et al.}~\cite{OSB22}, which conjectures that the second largest eigenvalue $\lmd$ of the transition matrix $R$ satisfies $({\rm Im}\lmd)^2/{\rm Re}\lmd\leq \dsgm$.
To compare our results, we introduce the corresponding eigenvector $\bsv$ normalized as $\sum_i \abs{v_i}^2/\pss_i=1$.
By setting $a_i={\rm Re}v_i/\pss_i$ and $b_i={\rm Im}v_i/\pss_i$ and comparing the relation $\lmd \bsv^\dagger \bsv=\bsv^\dagger R\bsv$, the oscillation fluctuation and the autocorrelation become equivalent to the real and the imaginary part of the second largest eigenvalue: ${\rm Im}\lmd=\alpha_{ab}$ and ${\rm Re}\lmd=D_a+D_b$~\cite{OIK23}.
However, to proceed to the above conjecture or a similar relation from our bounds, we need to evaluate the eigenvector $\bsv$, which appears in the form of $\|a^2+b^2\|$ in case of \eref{main-1}.
At present, we do not have good tools to examine this eigenvector in detail, which is left as a future problem.

We finally compare our results to the relations shown in Ohga. {\it et al.}~\cite{OIK23}.
In both inequalities, the fluctuation oscillation relative to the autocorrelation is bounded from above.
The difference lies in the fact that the bound shown in Ref.~\cite{OIK23} employs the maximum cycle affinity as a thermodynamic quantity, while our bounds employ entropy production rate.
To compute cycle affinity, we need detailed information on the system.
In contrast, the stationary entropy production rate depends only on heat currents, which can be measured through the total change in energy of baths.
In particular, \eref{main-1} connects oscillation fluctuation, autocorrelation, and entropy production rate directly.
We expect that the experimental verification of \eref{main-1} is tractable in several micro- and mesoscale stochastic systems and chemical systems.

\figin{8cm}{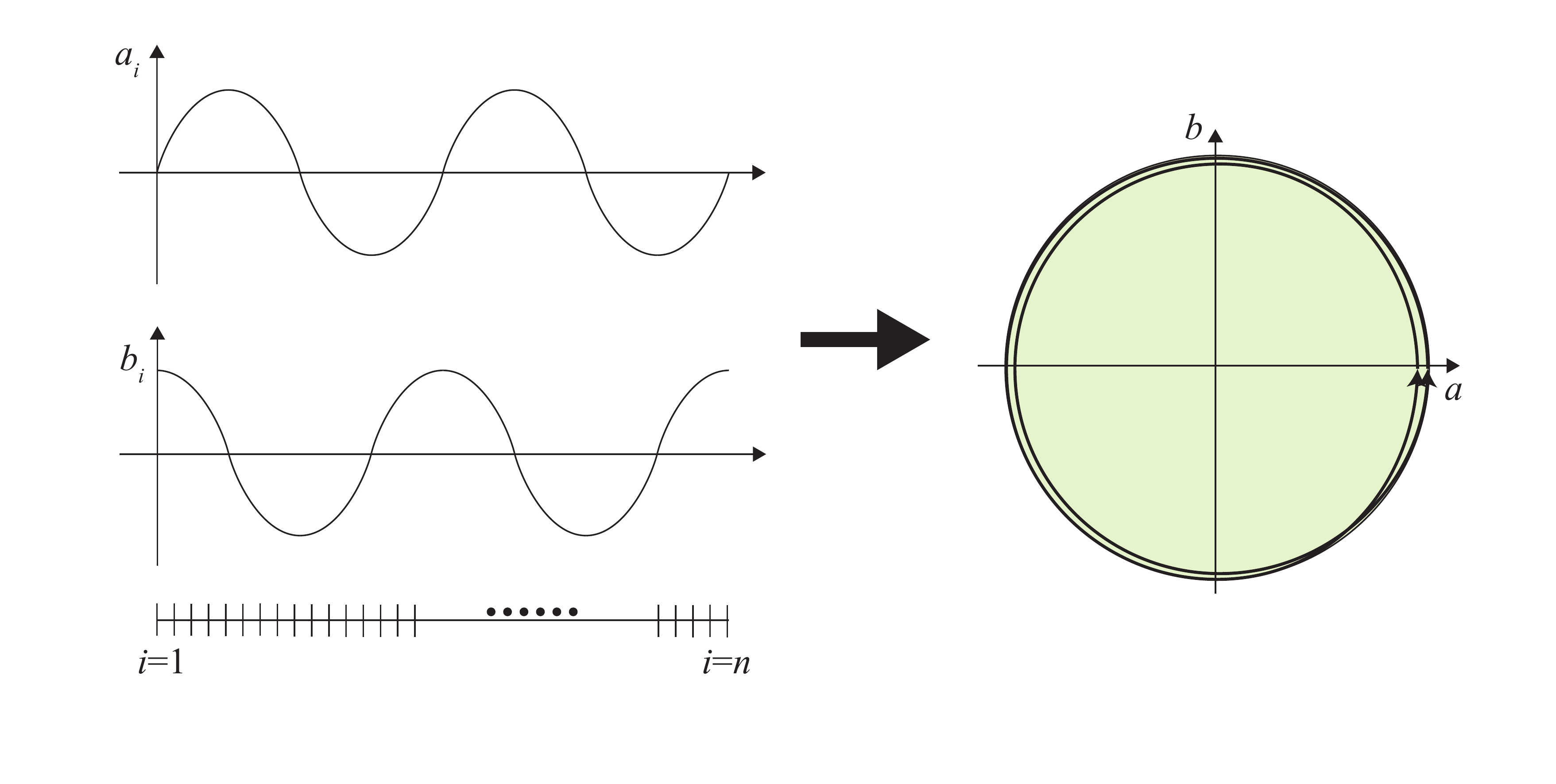}{
An example of $a_i$ and $b_i$ in a unicyclic system.
By taking $n\to \infty$ limit, the point $(a_i,b_i)$ moves on the circumference of the light green circle.
In this case, the point in $a$-$b$-space rotates twice per single period.
}{ex1}

\para{Example: unicyclic system}
To shed light on the power of our inequalities, we apply them to the simplest setup: a uniform unicyclic system with $n$ states, which is analyzed in Ref.~\cite{OSB22}.
We set transition rates as $R_{i+1,i}=ke^{\beta F/n}$ and $R_{i,i+1}=k$, where we identify state $n+1$ with state 1.
The stationary entropy production rate is computed as 
\eq{
\dsgm=k(e^{\beta F/n}-1)\frac{\beta F}{n}.
}

We set two observables as $a_i=\sin \omega i$ and $b_i=\cos \omega i$ with $\omega=2\pi m/n$, where $m$ is a natural number (The case of $m=2$ is drawn in \fref{ex1}).
Then, both the fluctuation oscillation and the autocorrelation are calculated as $\alpha_{ab}=\frac12 k(e^{\beta F/n}-1)\sin \omega$ and $D_a=D_b=k(e^{\beta F/n}+1)\sin^2 (\omega/2)$, whose ratio behaves as
\eqa{
\frac{2\abs{\alpha_{ab}}}{D_a+D_b}=\frac{\sin \omega \tanh \frac{\beta F}{2n}}{2\sin^2 (\omega/2)}\simeq \frac{\beta F}{2\pi m},
}{ex-lhs}
where the last approximation is valid under the large $n$ situation.
In addition, with noting $\|a^2+b^2\|=1$ and $\ell_C=2n\sin (\pi m/n)$, two definitions of a characteristic maximum speed of oscillation are calculated as $\wos=\abs{\alpha_{ab}}/\pi=\frac12 k(e^{\beta F/n}-1)\sin(2\pi m/n)/\pi$ and $\vos=4\pi \abs{\alpha_{ab}}/\ell_C^2=\frac12 k(e^{\beta F/n}-1)\pi \sin(2\pi m/n)/n^2\sin^2 (\pi m/n)$.
Thus, right-hand sides of two inequalities \eqref{main-1} and \eqref{main-2} read
\balign{
\frac{\dsgm}{2\pi \wos}&=\frac{\beta F}{n\sin \frac{2\pi m}{n}}\simeq \frac{\beta F}{2\pi m}, \\
\frac{\dsgm}{2\pi \vos}&=\frac{\beta F n^2 \sin^2 \frac{\pi m}{n}}{\pi ^2 \sin \frac{2\pi m}{n}}\simeq \frac{\beta F m}{2\pi}.
}
Comparing \eref{ex-lhs}, we see that \eref{main-1} achieves its equality for any $m$, while \eref{main-2} does only when $m=1$ and it is a loose bound by $m^2$ for $m\geq 2$.
This difference comes from the looseness of the isoperimetric inequality when the winding number of the poly-lateral around the center is not one.

\figin{6cm}{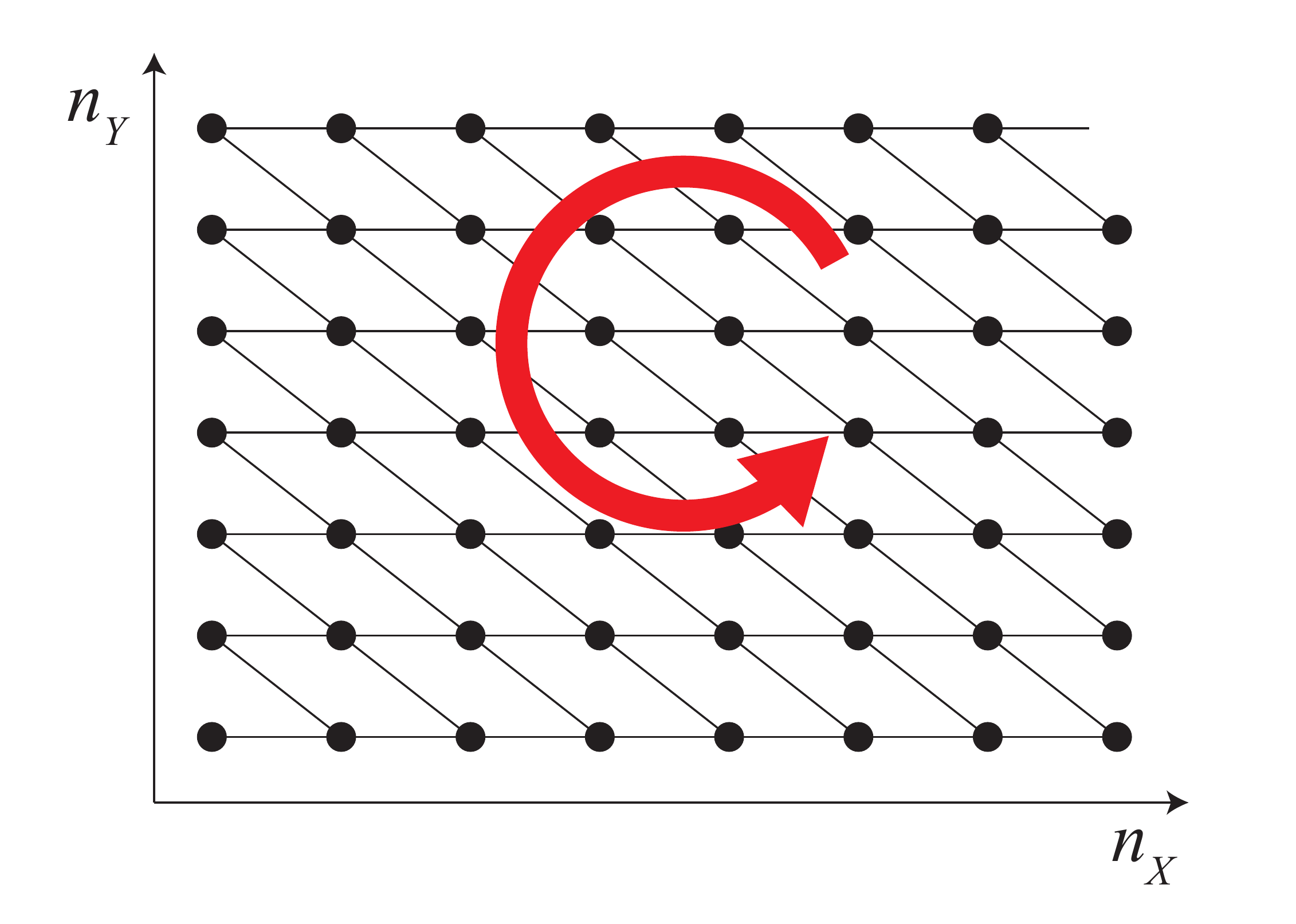}{
An example of the state space of a chemical reaction model, reversible Brusselator model: $A\lr X$, $2X+Y\lr 3X$, $B+X\lr Y+D$, $X\lr E$.
A single state is represented by a single vertex $(n_X, n_Y)$, and a state can jump to another state connected by an edge in a single transition.
We can expect oscillation (rotational flow) as the red arrow in some parameter regime, which can be captured by our bound \eqref{main-1} with a proper cutoff.
}{chemi}

\para{Other possible applications}
We here briefly draw other possible applications.
One important application is to two-dimensional Langevin systems in a confined region with rotational force.
This is straightforward by following a standard method~\cite{SST16, Shibook} that we first discretize the space and then take the continuous limit.
We note that quantities in \eref{main-1}, fluctuation oscillation, autocorrelation, entropy production rate, and the norm of $a^2+b^2$, do not diverge in this limit.

Another important application is to chemical reaction systems.
Some chemical systems including the Brusselator model show a nonequilibrium phase transition to a coherent oscillation phase~\cite{GPbook}.
From a microscopic perspective, the state of the system is a pair of particle numbers, and transition rates between two states are given from a chemical reaction network.
If a chemical system has two species $X$ and $Y$, for example, the microscopic state is given by $(n_X, n_Y)$.
An example of the state space (a reversible Brusselator model) is shown in \fref{chemi}.
A proper macroscopic limit recovers its deterministic time evolution.
Setting $a=n_X$ and $b=n_Y$, we can examine the magnitude of oscillation in $n_X$-$n_Y$ plane in terms of autocorrelation, entropy production rate, and the maximum number of species, which is well-defined under a proper cutoff.

\para{Proofs of Eqs.~\eqref{main-1} and \eqref{main-2}}
We derive two inequalities with replacing the entropy production rate $\dsgm$ by {\it pseudo entropy production rate}~\cite{Shi21, Shibook} 
\eq{
\dPi:=\sum_{(i,j)}\frac{(R_{ij}\pss_j-R_{ji}\pss_i)^2}{R_{ij}\pss_j+R_{ji}\pss_i}=\sum_{(i,j)}\frac{J_{ij}^2}{A_{ij}}.
}
Here, we defined the local activity, or traffic, in the stationary state as $A_{ij}=R_{ij}\pss_j+R_{ji}\pss_i$, which quantifies the frequency of jumps between $i$ and $j$.
Noting $\dPi\leq \dsgm$, we confirm that proving inequalities with $\dPi$ suffices for our purpose.

We employ a geometric interpretation with $a$-$b$ plane (see \fref{area}).
We introduce two distances, one is between two states $i$ and $j$ and the other is from the origin, as $\ell_{ij}:=\sqrt{(a_i-a_j)^2+(b_i-b_j)^2}$ and $r_i:=\sqrt{a_i^2+b_i^2}$, respectively.
The oriented area of the triangle with $i$, $j$ and the origin, with edges $\ell_{ij}$, $r_i$, and $r_j$, is expressed as $S_{ij}:=\frac12(a_jb_i-a_ib_j)$.
Using these quantities, the averaged autocorrelation $(D_a+D_b)/2$, which appears on the left-hand side of the main results as its denominator, is written as
\eq{
\frac{D_a+D_b}{2}=\frac12\sum_{(i,j)}A_{ij}\ell_{ij}^2.
}

A key fact to derive \eref{main-1} is that the oriented area of a triangle is always less than half of the product of two edges:
\eq{
S_{ij}\leq \frac12 \ell_{ij}r_i\leq \frac12 \ell_{ij} r_{\rm max},
}
where $r_{\rm max}:=\max_i r_i=\|\sqrt{a_i^2+b_i^2}\|$ is the maximum distance of point $(a_i,b_i)$ from the origin.
Using this relation and the Schwarz inequality, we have \eref{main-1}:
\balign{
\frac{2\abs{\alpha_{ab}^2}}{D_a+D_b}=&\frac{2\( \sum_{(i,j)}J_{ij}S_{ij}\)^2}{\sum_{(i,j)}A_{ij}\ell_{ij}^2}\leq \frac{r_{\rm max}^2}{2}\frac{\( \sum_{(i,j)}J_{ij}\ell_{ij}\)^2}{\sum_{(i,j)}A_{ij}\ell_{ij}^2} \nt \\
\leq& \frac{r_{\rm max}^2}{2}\sum_{(i,j)}\frac{J_{ij}^2}{A_{ij}}=\frac{r_{\rm max}^2}{2}\dPi\leq \frac{r_{\rm max}^2}{2}\dsgm. \lb{derive-1}
}
Here, we set the direction of edge $ij$ such that $J_{ij}$ is nonnegative.

We next derive \eref{main-2}, which requires a more complicated evaluation.
We consider a uniform cycle decomposition of current $J$ with cycle set $\calC$.
With this decomposition, we can set the direction of all cycles such that the current $J_C$ with any cycle $C$ is nonnegative.
We denote the $n$-th state in cycle $C$ by $i_n^C$, and define the {\it length} of cycle $C$ as $\ell_C:=\sum_{n=1}^{N_C} \ell_{i_{n+1}^Ci_n^C}$.
Here, $N_C$ is the number of states in cycle $C$ and we identify $i_{N_C+1}^C=i_1^C$.

Now, we employ the discrete isoperimetric inequality~\cite{FTT55, OIK23}.
The discrete isoperimetric inequality for $N_C$-lateral in the $a$-$b$ plane with $n$-th vertex $(a_{i_n^C}, b_{i_n^C})$ reads
\eq{
\( 4N_C\tan \frac{\pi}{N_C}\) \abs{\sum_{n=1}^{N_C} S_{i_{n+1}^Ci_n^C}}\leq \ell_C ^2.
}
Employing this relation, the fluctuation oscillation is evaluated as
\balign{
\alpha_{ab}=&\sum_C J_C \sum_{n=1}^{N_C} S_{i_{n+1}^Ci_n^C}\leq \sum_C \frac{J_C}{4N_C\tan \frac{\pi}{N_C}} \ell_C ^2 \nt \\
\leq&\frac{\max_C \ell_C}{4\pi}\sum_C J_C\ell_C=\frac{\max_C \ell_C}{4\pi}\sum_{(i,j)} J_{ij}\ell_{ij},
}
where we used $a \tan \frac \pi a\geq \pi$ for $0<a<\frac12$.
Following a similar transformation to \eref{derive-1}, we arrive at \eref{main-2}.

\para{Discussion}
We derived thermodynamic bounds on fluctuation oscillation in a simple form, which is easy to address experimentally.
Our result has wide applicability, from particle systems in continuous space, macroscopic systems, and chemical reaction systems, which is another advantage of our bound.

One may hope that the operator norm $\|\sqrt{a^2+b^2}\|$ in \eref{main-1} can be replaced by the stationary average $\langle \sqrt{a^2+b^2}\rangle$ or a similar quantity.
However, adopting an approach similar to ours, it appears not easy to derive such a relation.
In the left-hand side of \eref{main-1}, both the numerator ($\alpha_{ab}$) and the denominator ($D_a$ and $D_b$) employ the information of the stationary distribution $\pss$.
This is also true for the right-hand side:
Both the numerator ($\dsgm$) and the denominator ($\alpha_{ab}$) employ the information of the stationary distribution $\pss$.
Thus, there is no room for other quantities including $\|a^2+b^2\|$ to employ the information of the stationary distribution $\pss$.

\bigskip

\para{Acknowledgement}
The author is supported by JSPS KAKENHI Grants-in-Aid for Early-Career Scientists Grant Number JP19K14615.

\end{document}